\documentclass[%
preprint,
 amsmath,amssymb,
 aps,
]{revtex4-2}

\usepackage{epstopdf} 

\usepackage{graphicx}

\usepackage{dcolumn}
\usepackage{bm}
\usepackage{hyperref}

\usepackage{epstopdf}
\usepackage{lineno}
\usepackage{textcomp}
\usepackage[gen]{eurosym}
\usepackage{hyperref} 
\usepackage{slashed}
\usepackage[dvipsnames]{xcolor}  
\usepackage{placeins}

\usepackage{fancyheadings}
\usepackage{lastpage}
\usepackage{caption} 
\usepackage{verbatim}
\usepackage{color} 
\usepackage{subcaption} 
\usepackage{datetime}
\usepackage{booktabs, tabularx, multirow, array} 
\usepackage[separate-uncertainty,retain-explicit-plus,per-mode=symbol,binary-units]{siunitx}
\usepackage[version=4]{mhchem}
\DeclareSIUnit\c{\mbox{$c$}}
\DeclareSIUnit\year{yr}

\newcommand\snowmass{\begin{center}\rule[-0.2in]{\hsize}{0.01in}\\\rule{\hsize}{0.01in}\\
\vskip 0.1in Submitted to the  Proceedings of the US Community Study\\ 
on the Future of Particle Physics (Snowmass 2021)\\ 
\rule{\hsize}{0.01in}\\\rule[+0.2in]{\hsize}{0.01in} \end{center}}

\begin{document}

\snowmass


\title{ \textbf{ {\LARGE Introduction to a  low-mass dark matter project, \textcolor{DarkOrchid}{ALETHEIA}: \textcolor{DarkOrchid}{A} \textcolor{DarkOrchid}{L}iquid h\textcolor{DarkOrchid}{E}lium \textcolor{DarkOrchid}{T}ime projection c\textcolor{DarkOrchid}{H}amb\textcolor{DarkOrchid}{E}r \textcolor{DarkOrchid}{I}n d\textcolor{DarkOrchid}{A}rk matter} } }

\thanks{Corresponding authors: Junhui Liao and Yuanning Gao.}%

\author{Junhui Liao}
\email{junhui\_liao@brown.edu, junhui\_private@163.com}
\affiliation{
Department of Physics , Brown University
}
\affiliation{%
Division of Nuclear Physics, China Institute of Atomic Energy, Beijing, China 
}%

\author{Yuanning Gao}
\email{yuanning.gao@pku.edu.cn}
\affiliation{%
 School of Physics, Peking University 
}%

\author{Zhuo Liang, Zhaohua Peng, Jian Zheng, Jiangfeng Zhou}
\affiliation{%
The Division of Nuclear Physics , China Institute of Atomic Energy, Beijing, China 
}%

\author{Lifeng Zhang, Lei Zhang}
\affiliation{%
The Division of Nuclear Synthesis Technology, China Institute of Atomic Energy, Beijing, China 
}%

\author{Zebang Ouyang}
\affiliation{%
The School of Nuclear Technology, University of South China, Hengyang, Hunan, China 
}%


\date{\today}

\begin{abstract}
Dark Matter (DM) is one of the most critical questions to be understood and answered in fundamental physics today. Plenty of astronomical and cosmological observations have already pinned down that DM exists in the Universe, the Milky Way, and the Solar System. However, understanding DM with the language of elementary physics is still in progress. DM direct detection tests the interactive cross-section between galactic DM particles and an underground detector's nucleons. WIMPs is the most discussed DM candidate. After decades of hunting, a convincing WIMPs signal is still at large. Relatively, the low-mass WIMPs region ($\sim$ 10 MeV/c$^2$ - 10 GeV/c$^2$) has not been fully exploited compared to high-mass WIMPs ($\sim$ 10 GeV/c$^2$ - 10 TeV/c$^2$). By filling the arguably cleanest bulk material, LHe, into the arguably most competitive detector in the field, TPCs, ALETHEIA is supposed to achieve an extremely low-level background; therefore, to help answer one of the most pressing physical questions today: the nature of DM. In this paper, we briefly go through the physics motivation of low-mass DM, the ALETHEIA detector's design, possible analysis channels available for DM searches, and the progress we have made since the project launched in the summer of 2020.
\end{abstract}

\maketitle

\tableofcontents

\section*{The existence of dark matter and its detection} 

Astronomical evidence of many types, including cluster and galaxy rotation curves~\cite{Zwicky33, Rubin70}, lensing studies and spectacular observations of galaxy cluster collisions~\cite{Refreiger03,Clowe06,Fields08}, and cosmic microwave background (CMB) measurements~\cite{Planck2018ResultsOne}, all point to the existence of cold dark matter (CDM) particles.  Cosmological simulations based on the CDM model have been remarkably successful at predicting the actual structures we see in the Universe.  Alternative explanations involving modification of general relativity have not been able to explain this large body of evidence across all length scales~\cite{feng2014planning}. 
Recent results from Gaia~\cite{GaiaExp} show consistency with many previous experiments therefore much more securely pinned down than ever on: (a) DM dominates the mass of the Milky Way Galaxy~\cite{PostiHelmi19}, and (b) the local DM mass density in the Solar system is around 0.3 GeV/c$^2$ $\cdot$ cm$^{-3}$~\cite{HagenHelmi18}.

Weakly Interactive Massive Particles (WIMPs) is one of the most prominent  DM candidates.  WIMPs are a hypothesized class of DM particles that would freeze out of thermal equilibrium in the early Universe with a relic density that matches observations~\cite{Feng10}.  The so-called ``WIMP miracle'' is the coincident emergence of WIMPs with similar characteristics both from the solution of the gauge hierarchy problem and from the observed relic density of dark matter.

There are several viable strategies to detect DM.  Indirect detection experiments aim to observe high-energy particles resulting from the self-annihilation of DM.  Collider experiments look for the production of DM particles in high-energy collisions.  Direct detection experiments aim to observe the rare scatters of DM on the low threshold, very low background detectors operated in deep underground Laboratories. Table~\ref{tabDMExpCat} categorizes (most of) currently active DM experiments. The classification is not very strict in the sense that some experiments can have more than one categories in the table~\ref{tabDMExpCat}, or even beyond the scope of DM search. Taking LZ as an example, LZ is a high-mass DM experiment, but its physics searches are extensive, including but are not limited to: annual modulation, low-mass WIMPs search via S2O (S2 Only) analysis, Axion-like particles, two neutrinos double beta decay, and neutrinoless double beta decay, etc. Other leading DM experiments have a similar feature.

\newcommand{\otoprule}{\midrule[\heavyrulewidth]} 

\begin{table}[ht]
   \centering
   \caption{Categorization of DM experiments with physics motivation} \label{tabDMExpCat}   
      \begin{tabular}{c c c c}
      \toprule%
         \multicolumn{1}{c}{\bfseries{ low-mass } }  &
         \multicolumn{1}{c}{\bfseries{ high-mass} }  &
         \multicolumn{1}{c}{\bfseries{Annual modulation} }  &
         \multicolumn{1}{c}{\bfseries{Directional} }     \\%
	  \otoprule%
      ALETHEIA, CEDX,				&ArDM, DarkSide, 		&ANAIS,				& DRIFT, \\
      CRESST, DAMIC, Edelweiss,		&DEAP, LZ, PandaX, 	&COSINE-100, 		& NEWAGE, \\
      SENSI, SuperCDMS.				&PICO, XENON.	 	&DAMA/LIBRA.			& NEWS. \\
      \bottomrule
    \end{tabular}
\end{table}

This paper will focus on the \textbf{ALETHEIA} project for the experimental detection and verification of the WIMPs hypothesis.

\subsection*{The context of  low-mass WIMPs direct detection}

The lowest limit for WIMPs-nucleon interaction for $\sim$ 30 GeV/c$^2$ is down to $\sim$ 10$^{-48}$ cm$^2$~\cite{Xenon1TPaper18, PandaX2021, LZ2022}, which are roughly nine and two orders lower than the interaction hypothetically mediated through the Z boson ($\sim$ 10$^{-39}$ cm$^2$) and the Higgs boson ($\sim$ 10$^{-46}$ cm$^2$), respectively. However, in theory, the WIMPs scenario could still be ``long-life'', as mentioned in reference~\cite{HooperBrownTalk2018}. Among other possibilities, one way to reconcile the conflict, WIMPs are supposed to be detected but failed, is that dark matter might turn out to be lighter than the mass region where  high-mass experiments (DarkSide, LUX/LZ, PandaX, and Xenon) can effectively reach. For instance, $\sim$ 100s MeV/c$^2$ - 10 GeV/c$^2$. In this note below, unless specially mentioned, we will consider $\sim$ 100s MeV/c$^2$ - 10 GeV/c$^2$  as a default  for low-mass DM. 

In fact, independent of SuperSymmetry (SUSY) models which predict weakly mass WIMPs~\cite{Jungman96}, there are four categories of well-motivated scenarios favor MeV/c$^2$ - GeVc/c$^2$ dark matter.

(a) The ``WIMPs miracle'' model motivates $\sim$ 10 MeV/c$^2$ - 100 TeV/c$^2$ WIMPs~\cite{Feng10, FengKumar08}: a stable particle with such mass annihilating each other with a cross-section of $\sigma v \sim 2 \times 10^{-26} \text{cm}^3$/s in the early university would result in consistent dark matter density as measured~\cite{Steigman12}. DM mass greater than 240 (340) TeV/c$^2$  would violate the requirement of partial wave unitarity assuming DM is a Dirac (Majorana) fermion~\cite{Griest90}, while the Big Bang nucleosynthesis would ruin if an annihilating relic with a mass lighter than $\sim$ 1-10 MeV/c$^2$~\cite{Bohm13}.

(b) Light DM annihilates to quarks might have been suppressed during the epoch of recombination; instead of coupling to quarks, light DM would possibly couple to leptons, in particular, electrons. In direct detection, DM could scatter with electrons to generate individual electrons, individual photons, individual ions, and heat/phonons~\cite{Essig12}. 

(c) The asymmetry in the dark sector might be related to the baryon asymmetry (matter and anti-matter),  resulting in zero net baryon number of the universe ~\cite{Cohen93, Zurek14}. The mass of DM would be $\sim r \times (4-5)$ GeV/c$^2$, where $r$ is a factor that maintains equilibrium between the dark and SM (Stand Model) sectors in the early universe. If $r = 1$, the DM mass would be $\sim (4-5)$ GeV/c$^2$. 

(d), Strongly Interacting Massive particle (SIMP) models propose dark matter as a meson- or baryon-like bound-state of hidden sector particles, with a mass near the QCD scale, $\sim$ 100 MeV/c$^2$~\cite{StrasslerZurek07, Hochberg14, Hochberg15,Kuflik16,Kuflik17}. Considering constraints from the CMB, dark matter masses might be the range of $\sim$ 5 - 200 MeV/c$^2$. 

Regardless of whether or not WIMPs could be discovered at the scale of $\sim$ 100 GeV/c$^2$, low-mass WIMPs should still be investigated. Since (1) it has been motivated by several different mechanisms, as mentioned above; (2) considering there exist tens of elementary particles in the Standard Model, it is reasonable to guess the existence of more than one dark matter particle, which would naturally have different masses. So, even if high-mass WIMPs were discovered; physicists should also check the low-mass region to see if there is low-mass dark matter there and vice versa.
However, as mentioned, none of the leading high-mass DM experiments is sensitive to low-mass WIMPs due to their relative heavy target material, xenon, or argon~\footnote{I am focusing on the traditional Spin-Independent model with ``S1/S2'' analysis. By applying other analysis methods like S2O and new models like Migdal-effect, these liquid heavy noble gas experiments could be sensitive to WIMPs mass down to the sub-GeV region. We will address this in detail below.}, though these experiments could achieve extremely low backgrounds. 

The ALETHEIA was inspired by both high-mass and low-mass WIMPs experiments in the community. Although there existed quite a few low-mass DM experiments~\cite{SuperCDMS2014, SuperCDMS2016, DAMIC2016, CRESST2014, CRESST2017, EDELWEISS2017, CDEX2018}, we believe the ALETHEIA is a competitive project in the race of hunting for low-mass DM thanks to several advantages; for details, please refer to the section of ~\hyperref[secProjectSubWhyLHeTPC]{Why LHe TPC?}.

ALETHEIA is not the only project using LHe for DM searches. There are a couple of experiments aiming for DM hunting with Superfluid Helium (SHe), but with different readout methods. For instance, HeRald experiments~\cite{Hertel19, SpiceHeRald21} and reference~\cite{Maris17} search for $\sim$ MeV/c$^2$ DM with Transition Edge Sensor (TES) and ionization field, respectively. Helium gas becomes liquid as long as being cooled to 4.5 K; this is the ``general'' LHe. Keep cooling to 2.17 K or below, and LHe reaches the superfluid status. SHe has many different features than the general LHe. One of them is heat conductivity. As a good heat conductor, SHe can detect ``quasiparticles'' (phonons and rotons) left by incident particles. Thanks to the low energy of quasiparticle production, 0.62 meV, a SHe detector can, in principle, detect $\sim$ MeV/c$^2$ DM. However, the SHe detector must work at an extremely low temperature as 100 mK or below~\cite{Hertel19}\footnote{Actually, the temperature for other similar cryogenic detectors is $\sim$10 mK. For instance, the temperature of the CUORE and SuperCDMS detectors are, 10 mK~\cite{CUORE2022} and 15 mK~\cite{SuperCDMS2016}, respectively}.In terms of cryogenic engineering, it is more challenging to build a SHe detector than a general LHe one.

\section*{Introduction to the ALETHEIA experiment}
\begin{figure}[!t]	 
	\centering
    \includegraphics[width=6.5in, angle = 0]{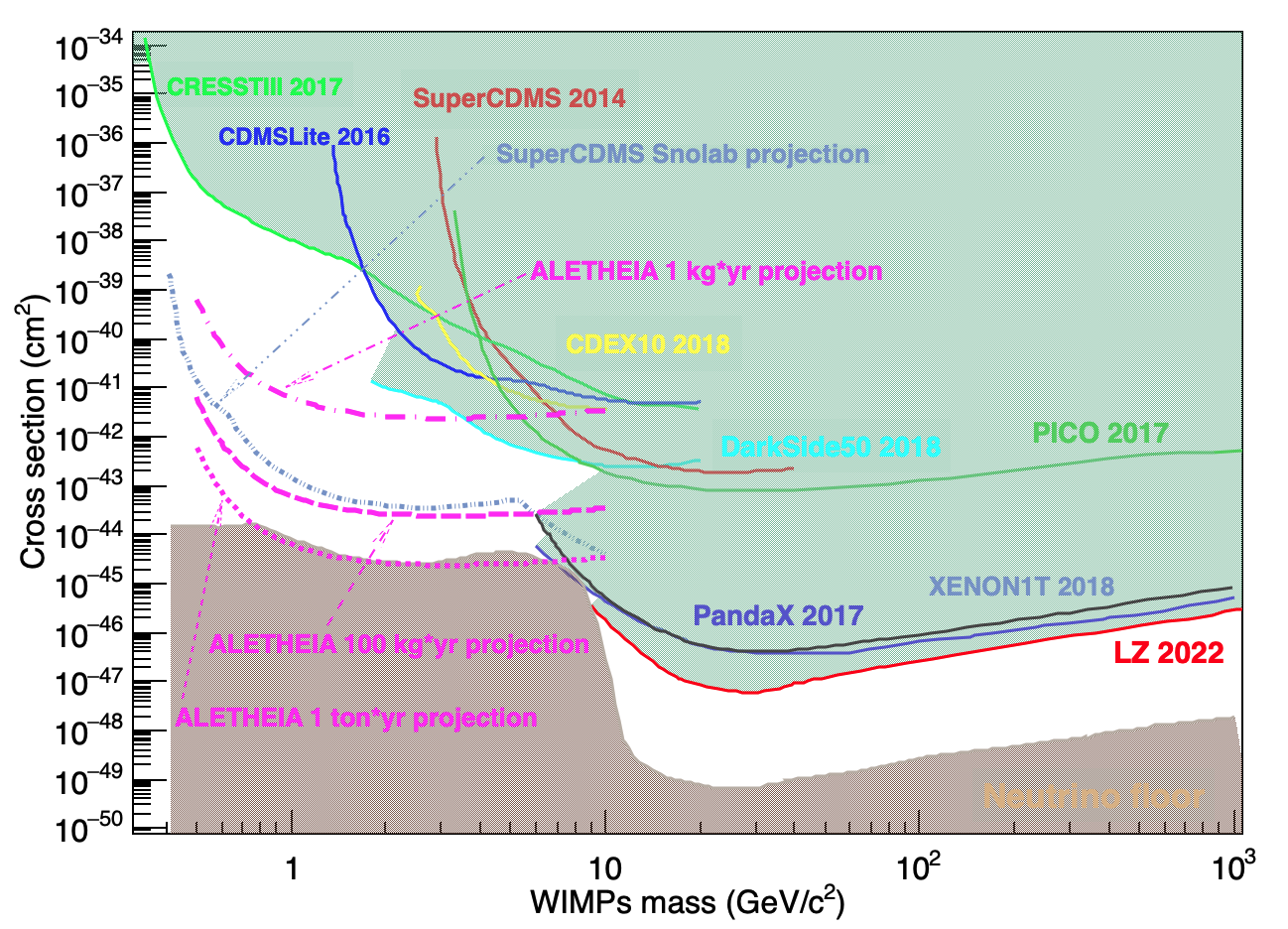}
	\caption{The figure shows the parameter space for the spin-independent WIMPs-nucleon cross-section. The dark green region represents the space that has been excluded by leading direct detection experiments at a 90\% confidence level. The area with light brown color corresponds to the space of the ``neutrino floor'' where neutrinos can generate WIMPs-like events in DM detectors. The projected sensitivities of the ALETHEIA with the exposure of 1 kg*yr, 100 kg*yr, and 1 ton*yr are shown. The upper limits or projected limits of other experiments are also show. The ``CRESSTIII-2017'' data cited from~\cite{CRESST2017}, ``CDMSLite-2016'' from~\cite{CDMSLite2016}, ``CDEX10 2018'' from~\cite{CDEX2018}, ``DarkSide50 2018'' from~\cite{DarkSide502018}, ``LZ 2022'' from \cite{LZ2022}, ``PandaX 2021'' from~\cite{PandaX2021}, ``PICO 2017'' from~\cite{PICO2017}, ``Neutrino floor'' from~\cite{Billard14}., ``SuperCDMS 2014'' from~\cite{SuperCDMS2014}, ``SuperCDMS Snolab projection'' from~\cite{SuperCDMS2016}, ``XENON1T 2018'' from~\cite{Xenon1TPaper18}.}\label{projectedSensitivities} 
\FloatBarrier
\end{figure}

The ALETHEIA project is an instrumental background-free experiment with the ROI (Research Of Interest) of $\sim$ 100s MeV/c$^2$ - 10 GeV/c$^2$. Fig~\ref{projectedSensitivities} shows the projected sensitivity of the ALETHEIA with the exposure of 1 kg*yr, 100 kg*yr, and 1 ton*yr, respectively. The projected sensitivity is based on two assumptions: (a) there is no background in the interesting recoil energy range (after a series of cuts being applied), 0.5 keV$_{nr}$ - 10 keV$_{nr}$; (b) the detection efficiency is 100\% for the whole ROI energy interval. We made such assumptions under the context of no data available. We will update the projections as long as we have data. 

With the S2 Only (S2O) analysis, the ALETHEIA project could be sensitive to WIMPs mass lower than 100s MeV/c$^2$. However, because the S2O analysis depends on data, it is impossible to project the sensitivity at this stage. 

Here, instrumental background-free means that in the range of ROI, a small number of background events (for instance, $<$ 0.1 events) are expected, such that zero background events are observed. The instrumental backgrounds include radioactive particles due to the materials in the detector system (including dust) and the particles generated by cosmological muons hitting the rocks near the detector or the detector itself. For low-mass WIMPs searches, another background is registered by $^8$B solar neutrinos. The $^8$B events can't be discriminated from low-mass WIMPs signals. According to reference~\cite{Bahcall04, Billard14, SNO02}, the measured $^8$B events are well consistent with the theoretical prediction. The uncertainty of the events is $\sim$16\%, as reported in reference~\cite{SNO02}. Under the context of detecting WIMPs at a cross-section below the $^8$B neutrino floor, if a detector that is free of instrumental backgrounds, by subtracting (theoretically estimated) $^8$B events from the observed events, it is still possible to to detect WIMPs directly at a underground lab.\\

\section*{The design of the ALETHEIA detector}
The design of the dual-phase ALETHEIA has been significantly inspired by two communities: LHe and DM direct detection. In the LHe community, efforts have been made for DM searches. The HERON (HElium Roton Observation of Neutrinos) project aims at creating a superfluid helium detector to detect Solar neutrinos~\cite{HOERN1}. The researchers also considered using it for dark matter hunting~\cite{HOERN88, HOERN13}. Unlike these earlier concepts, the ALETHEIA project will use liquid helium at $\sim$ 4 K, above the Lambda curve, instead of superfluid helium. Reference~\cite{GuoMckinsey13} also proposed and simulated a liquid helium TPC hunting for DM in 2013 with ``S1/S2'' analysis, where ``S1'' refers to prompt scintillation while ``S2'' is the electroluminescence light originated from ionization by recoiled target nuclei. 
The ALETHEIA has also referred to many other successful experiments which implemented TPCs, such as DarkSide, LUX/LZ, PandaX, XENON, etc.

 \begin{figure}[!t]	 
	\centering
    \includegraphics[width=5.0in, angle = 0]{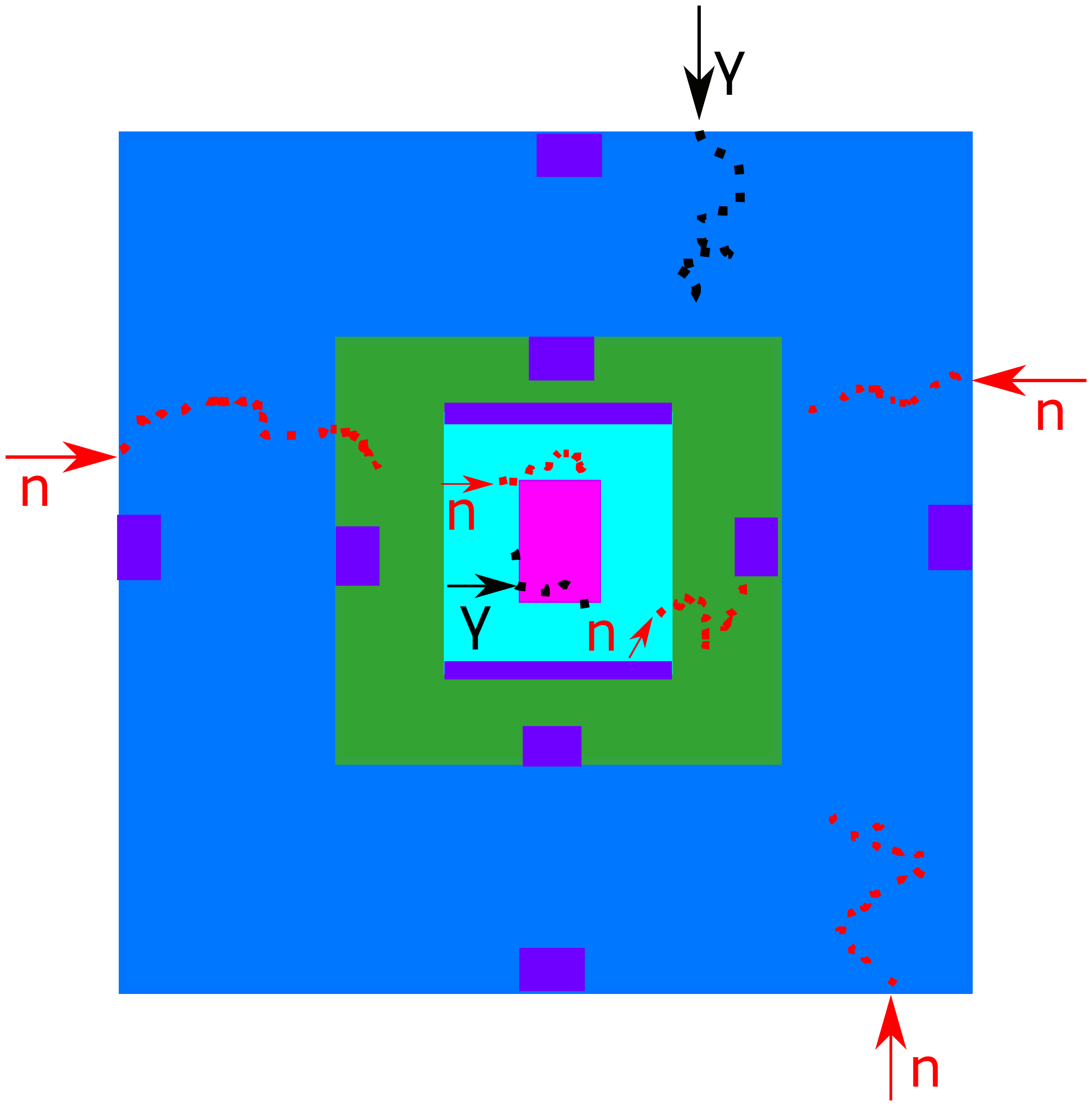}  
	\caption{The schematic drawing of the ALETHEIA detector (not to scale). From outside to inside: The light blue area represents the water tank surrounding the whole detector system, with a diameter of a few meters; the four purple rectangles on the edge of the water tank are the PMTs to detect background signals insides of the water tank; the dark green is the Gd-doped liquid scintillator veto, with a thickness of $\sim$ half-meter; the four purple rectangles at the edge of veto (green area) are PMTs to detect signals insides of it; the cyan area is the active volume of the TPC, filled with liquid helium; the two horizontal dark blue stripes on the top and bottom of the active volume represent the SiPMs to detect scintillation and electroluminescence; the magenta region represents the fiducial volume of the TPC where extremely low backgrounds are expected there. Red dots represent background neutrons. Black dots are background $\gamma$s.}
	\label{FigSchematicALETHEIA}
\FloatBarrier
\end{figure}

 Figure~\ref{FigSchematicALETHEIA} shows the schematic drawing of such a detector, the core of the ALETHEIA experiment is a dual-phase liquid helium TPC (in cyan). The TPC center is magenta, representing the fiducial volume where extremely low or zero background is expected. On the top and bottom of the TPC are SiPMs (purple). The TPC was surrounded by a Gd-doped scintillator detector, which acts as a veto. The outmost is a water tank with a few meters diameter to shield neutrons and gammas outside of the detector system. 

For neutrons that come from outside the water tank, a few meters of water should be thick enough to thermalize them. The $\sim$ half-meter Gd-doped liquid scintillator would capture thermalized neutrons. For neutrons originating from the TPC inside, it could be identified by the feature of multiple hits in the TPC and (or) liquid scintillator because the interaction between neutron and detector materials is strong interaction. A typical WIMPs signal would only have one hit registered due to a much lower coupling between WIMPs and helium nuclei than the strong interaction. For $\gamma$s from outside of the water, the water tank can block them from entering the central detector; for $\gamma$s from inside of the TPC, ``S1/S2'', or PSD (Pulse Shape Discrimination), or a hybrid analysis combined both could discriminate from nuclear recoils induced by a neutron or a WIMPs.

\section*{Why LHe TPC?}\label{secProjectSubWhyLHeTPC}

An LHe TPC has several unique advantages, enabling the ALETHEIA detector to achieve a background-free or extremely low backgrounds search.

(a) High recoil energy. Helium is the lightest noble element and the second light element. Therefore, the same kinetic energy of incident WIMPs would result in greater recoil energy than other elements except for hydrogen. Although hydrogen is even lighter, hydrogen is not a good detector material due to a few drawbacks such as explosive, chemically active, etc. 

(b) High QF (Quenching Factor). For 16 keV nuclear recoil energy, the measured QF of LHe is $\sim$ 65\%~\cite{Santos08}; while LAr is $\sim$ 24\%~\cite{QFArgon14}, which is a factor of 3 smaller. Another measured QF of helium at 1.5 keV recoil energy is up to 22\%~\cite{Santos08}. As a comparison, the measured QF of hydrogen at 100 keV is only 2\%~\cite{Reichhart12}, and the estimated QF at 1.5 keV nuclear recoils would be much lower than 2\%. As a result, the QF of hydrogen is guaranteed to be at least one order smaller than helium at $\sim$1 keV$_{\text{nr}}$. Thus, for the same incident kinematic energy, although hydrogen has 4-fold greater recoil energy than helium, by considering their quenching factors, the electron equivalent recoil energy of hydrogen is at least a factor of 2.5 smaller than helium. 

One should also be aware that in the QF-measurement paper~\cite{Santos08}, only ionization is registered with a Micromegas detector, and excitation-induced scintillation is ignored, which is different from our detector. In our case, an LHe TPC can collect scintillation produced both by excitation and ionization. Consequently, the scintillation yield in an LHe TPC should be greater than the quenched energy alone since the process of excitation and de-excitation can also contribute scintillation.

(c) Radioactivity free. Helium has no radioactive isotopes; both $^3$He and $^4$He are stable elements. However, LAr and LXe have radioactive isotopes, and most (if not all) of these isotopes are very difficult to eliminate.

(d) Easy to purify. At 4 K temperature, $^3$He is the only solvable material in LHe~\footnote{Hydrogen can also solve in LHe, but the solubility is as low as 10$^{-14}$~\cite{JEWELL1979682} at 1 K. Hydrogen is not a radioactive element so it will not contribute background events.} while it is very rare, the ratio of $^4$He and $^3$He in Nature is $10^7$:1; any other impurities would exist in the solid-state and, therefore, easy to be purified with getters and cold-traps. LAr and LXe have solvable impurities and are not easy to get purified, although the purification level is continuously improved~\cite{XENONnTERAnalysis2022, XENONnT2022IDMTalk}. Actually, one of the most critical advantages of implementing liquid noble gas TPCs to hunt for DM is that they can purify their bulk material online and keep it extremely clean, achieving a very low background. As a comparison, solid detectors can not do similarly; this is part of the reason they are not competitive in terms of background constraints.

(e) Possible strong ER/NR discrimination. TPC technology is well-developed in LXe and LAr experiments and helped these projects to take a leading role in high-mass DM search in the past decade, one after another. These world-leading experiments demonstrated that the S1/S2 and/or PSD analysis could be implemented to achieve efficient ER/NR discrimination. As a TPC filled with another noble gas, helium; hopefully, ALETHEIA could mimic the success of peer experiments and help to explore the parameters space other experiments might be difficult to touch.

(f) Scalability. LXe and LAr experiments demonstrated that liquefied noble gas-filled TPCs could scale from $\sim$ kg up to $\sim$10 ton, even multi-10-ton. Therefore, we hope that an LHe TPC can also scale up to be a 1-ton and multi-ton detector, with which ALETHEIA could fully touch down the B-8 neutrino floor (neutrino fog~\cite{OHare21, NextGenerationLXeObservatory22}) or even below. 

And (g) Helium procurement. Helium is significantly cheaper than xenon. The price of helium is $\sim$ 1/7 of xenon (Though the prices of helium and xenon both fluctuate.). Every year, helium consumption is $\sim$ 20 k ton globally~\cite{heliumPriceUCSB}. One can buy 1-ton LHe without affecting the market price since it is only 0.005\% of the total consumption.\\

Although the reference~\cite{GuoMckinsey13} proposed the concept of an LHe TPC, no such detector has ever been built worldwide. As a creative technology, an LHe TPC naturally has some risks. To better understand the technical risks and challenges, we invited a few leading physicists in the field of LHe, TPC, and DM theory to review the ALETHEIA project in Oct 2019~\cite{DMWS2019PKU}. 

\section*{The review of the ALETHEIA project}

In Oct 2019, we organized a DM workshop at Peking University in Beijing, China~\cite{DMWS2019PKU}. Dr. J. Liao presented the project's concept during the workshop~\cite{LiaoPresentationDMWS2019PKU}~\footnote{The then name of the project is ALHET which stands for A Liquid HElium Time projection chamber.} and provided a $\sim$ 20-page documents~\cite{ALETHEIA_2019WorkPlan} to address the project in details. A panel composed of leading physicists~\footnote{The review panel member are: Prof. Rick Gaitskell at Brown, Prof. Dan Hooper at Fermilab at the University of Chicago, Dr. Jia Liu at the University of Chicago (now an assistant Prof. at Peking University.), Prof. Dan McKinsey at UC Berkeley, Dr. Takeyaso Ito at Los Alamos National Laboratory, and Prof. George Seidel at Brown University.} in DM and liquid helium reviewed the ALETHEIA project~\footnote{ALHET is the then name of the project. Its current name is ALETHEIA.}. The panel stated, ``\textit{It is possible that liquid helium (TPC) could enable especially low backgrounds because of its powerful combination of intrinsically low radioactivity, ease of purification, and charge/light discrimination capability}''~\cite{ALETHEIA_2019Review}. 
The panel also suggested two R\&D phases: 30 g and 10 kg LHe detectors. 
The 30 g phase represents a cell having a total mass of 30 g LHe, with a cylindrical shape, radius = 5 cm, height = 3 cm. There will be multiple versions of this setup to test: \\
Cal-I: ER and NR calibrations. \\
Cal-II: S2 signal optimizations. \\
Cal-III: SiPM testing at 4 K. 

The 30 g detector program is suited to answer the initial questions concerning the fundamental responses of the liquid helium to incident particles (neutrons and gammas/electrons) and establish the optimum conditions for operation. We have built a couple of 30 g detectors at CIAE, cooled the detectors down to the LHe temperature, and tested SiPM at 4.8 K. For details, please refer to the section of \hyperref[ALETHEIAprogress]{The progress of ALHETHEIA so far}.

The subsequent 10 kg system ($\sim$ 10 kg LHe, cylindrical shape, diameter = height = 45 cm.) would be necessary in order to demonstrate the viability of the elevated HV levels needed for an even larger scale dark matter search experiment. It would also test other needed aspects, such as large multi-channel photodetector arrays. Given the potentially slow drift speed of the ionization signals in an LHe TPC (Typical velocity of electron bubbles in LHe is $\sim$ 2 m/s.), to constrain the overlaps in background events due to cosmic rays, the 10 kg detector would be operated underground, instead of running above ground.

\section*{Implementing PSD, S1/S2, and S2O into the ALETHEIA for WIMPs search}
\subsection*{PSD for ALETHEIA}
Energetic particles passing through a medium of LHe will deposit part or all of its kinetic energy. If the incident particle is an electron (or $\gamma$), it interacts with the electrons of helium atoms electromagnetically. The atoms then get ionized or excited, or both.

If the incident particle is a neutron, it interacts with helium atoms strongly. The atoms get recoiled energy and become moving $\alpha$ particles. The $\alpha$ particles further interact with the electrons of surrounding helium atoms electromagnetically. Although electrons and $\alpha$ particles interact with helium atoms via electromagnetic interaction, the charge densities of ions and electrons are different. As mentioned in reference~\cite{McKinsey03}, for high-energy ($\sim$ MeV) electrons, energy deposition is 50 eV $\mu$m$^{-1}$; while for $\alpha$ particles, energy deposition is 2.5 $\times$10$^4$ eV $\mu$m$^{-1}$. For liquid xenon, similar results were obtained with simulation~\cite{DahlPhDThesis09}.

Essentially, a charge density depends on stopping power, or ``dE/dx''~\cite{McKinsey03, AprileDokeReview2009}. According to the \textit{Bethe formula}, for low energy incident particles ( $v \ll c$, where $v$ is the velocity of particles, $c$ is the speed of light.), $dE/dx \sim \propto 1/v$, while for the same kinetic energy $\alpha$s and electrons (100 keV for instance), the velocity of electrons is roughly 2 orders faster than $\alpha$s, as a result, $dE/dx$ of electrons is roughly 2 orders smaller than $\alpha$s. Consequently, the charge densities for ER/NR have two orders difference. Moreover, the geometry and separation of the tracks induced by ER and NR are different: ER events are ``small dots'' shape and separated on average 500 nm; NR events are the cylindrical shape and separated, on average, 1 nm. Considering the distance between the ion and its separated electrons is roughly 20 nm, ER events are well separated (500 nm $ \gg $ 20 nm), and the recombination is geminate~\cite{Onsager38}, meaning the recombined electron-ion pair is the one being separated moments ago. In comparison, NR events are heavily overlapped between individual ionization events (1 nm $ \ll $ 20 nm)~\cite{McKinsey03}, the recombination is columnar~\cite{Jaffe1913, Kramers52}, meaning the recombined electron-ion pair is not necessarily the one just separated. 

The PSD technique exploits the time feature of scintillation to discriminate ER/NR events, specifically, the lifetime of scintillation produced in singlet and triplet excimers: For LAr, the lifetime of singlet and triplet is a few ns and $\mu$s, respectively. The two orders difference in a lifetime is proved to be able to apply PSD for ER/NR in DarkSide-50 and DEAP. However, for LXe, the lifetime difference for singlet and triplet is only one order. Therefore, LXe experiments have not implemented PSD.

LAr has a fast component scintillation, $\sim$ 7 ns, which decays from excited singlets, and a slow component of 1.6 $\mu$s, which decays from excited triplet~\cite{Lippincott08}. However, LHe scintillation has three lifetimes, $< 10$ ns, 1.6 $\mu$s, and 13 s~\cite{McKinsey03}; the $< 10$ ns and 1.6 $\mu$s scintillation result from singlet decay, the 13 s due to triplet decay. According to the test with a few MeV incident $\alpha$ sources, the 1.6 $\mu$s component is weaker than the $< 10$ ns and 13 s~\cite{McKinsey03}. For the ALETHEIA's ROI energy, 0.5 - 10 keV$_{nr}$, the light yield of $\alpha$ and $\beta$ has not been measured directly. The relative cross-section of ionization and excitation is recoil-energy dependent: for $\sim$ MeV recoil energies, ionization is a factor of 5 greater than excitation; while for  $\sim$ keV$_{nr}$, the relative cross-section is flipped, the excitation is a factor of a few greater than ionization~\cite{ItoSeidel13}. As mentioned in reference~\cite{SeidelPKUTalk2019}, precision predictions about scintillation light yield in the various channels are uncertain for the moment (due to a lack of experimental data). In summary, only dedicated experiments can tell the viability of the PSD analysis on an LHe TPC.

For ER, with a 1.0 cm scale apparatus, reference~\cite{GuoJinst12} observed explicitly ~$\sim$ 10 ns and 1.6 $\mu$s scintillation components. Reference~\cite{Phan20} studied the effect of an electric field on scintillation yield.
Reference~\cite{Ito2012} studied scintillation produced by NR events with a 5 cm scale apparatus. An $^{241}$Am source with decayed 5.5 MeV $\alpha$ particles were used for the tests. 

\subsection*{S1/S2 for ALETHEIA}

LXe dark matter experiments have demonstrated that the S1/S2 technique can be utilized for discriminating ER/NR events and defining a fiducial volume (with S2). In principle, the same analysis methods could be transplanted into the ALETHEIA because the difference of charge density of ER and NR events in LXe also holds in LHe~\cite{McKinsey03}. 

The S2 signal resulted from dragging electrons away from the recombination process with an external electric field.  
Reference~\cite{Seidel14} indicates that ER and NR do not have the same fraction of elections collection for a certain electric field, which could be implemented in the S1/S2 analysis for events discrimination. Reference~\cite{GuoMckinsey13} showed a good ER / NR discrimination in LHe for ionization energy down to 10 keV$_{\text{ee}}$ with the drift field of 10 kV/cm, and photosensors can collect 20\% S1 scintillation.

The drift field of LUX is 170 V/cm (LZ is 310 V/cm, and DarkSide-20k is 200 V/cm.). While for an LHe TPC, the drift field is required to be 10 kV/cm or higher since the electron collection is only 40\% even under 10 kV/cm~\cite{Seidel14},  though the latest measurements~\cite{Phan20} showed a greater collection efficiency, $\sim$ 50\% at 10 kV/cm. 
One of the critical R\&D programs for the ALETHEIA project is developing a safe and stable HV system up to 500 kV or higher. 
Reference~\cite{SethumadhavanPhDThesis07} showed under the voltage of $\sim$ 500 V/cm, significant electroluminescence current can generate. This test demonstrated that a dual-phase TPC filled with helium could generate electroluminescence (or S2) signals, therefore, making the S1/S2 and S2O analysis of the ALETHEIA viable.

\subsection*{S2O for ALETHEIA}
The strategies of S2O analysis are not the same for DarkSide-50 and XENON-1T. For DarkSide-50, as mentioned in reference~\cite{DarkSide502018S2O}, to reach the lowest possible S2 signals, the fiducial volume can not be reconstructed with usual algorithm due to low photoelectron statistics (for S2), the fiducial region for the S2O analysis is in the $x-y$ plane by only accepting events where the largest S2 signal is recorded in one of the seven central top-array PMTs. For XENON-1T~\cite{XENON1TS2O19}, they used 30\% of Science Run (SR1) data as training data to determine events selections. Limits settings are computed using only the remaining 70\% data, which was not examined until the analysis was fixed. 
The critical feature to convince us that the S2O analysis makes sense is that the observed backgrounds are consistent with anticipated events for selected fiducial volume or datasets.
The S2O analysis on the ALETHEIA can not be decided without data on hands. Therefore, we will choose the most appropriate analysis once we have scientific data.

\section*{The progress of ALHETHEIA so far}\label{ALETHEIAprogress}

\subsection*{The progress of a 30 g LHe prototype detector}

We have designed our first 30 g LHe cell at CIAE in Beijing, China, as shown in references~\cite{TPBCoutingALETHEIA22, ALETHEIA-EPJ-22}. The primary purpose of this detector is to gain experience in building an apparatus capable of working at LHe temperature ($\sim$ 4.5 K), make sure external DC High Voltage (HV) could be applied to the detector, and the dark current of the detector under HV should be small enough.

To cool the 30 g detector down to 4.5 K, one must first check the vacuum of the whole detector system is good enough or not; if not, the vacuum-leaking places would lead to cryogenic leakage either. Consequently, the detector can not reach the LHe temperature. Even if the design of the detector is perfect, vacuum leakage can still happen anywhere due to mechanical failures; some of them are difficult to track.

After many efforts have been made (Please read reference~\cite{ALETHEIA21} for details), we ultimately cooled our detector down to LHe temperature in summer 2021, as shown in reference~\cite{TPBCoutingALETHEIA22, ALETHEIA-EPJ-22}.

\begin{figure}	
	\centering
	\begin{subfigure}[t]{3.1in}
		\centering
		\includegraphics[scale=0.39]{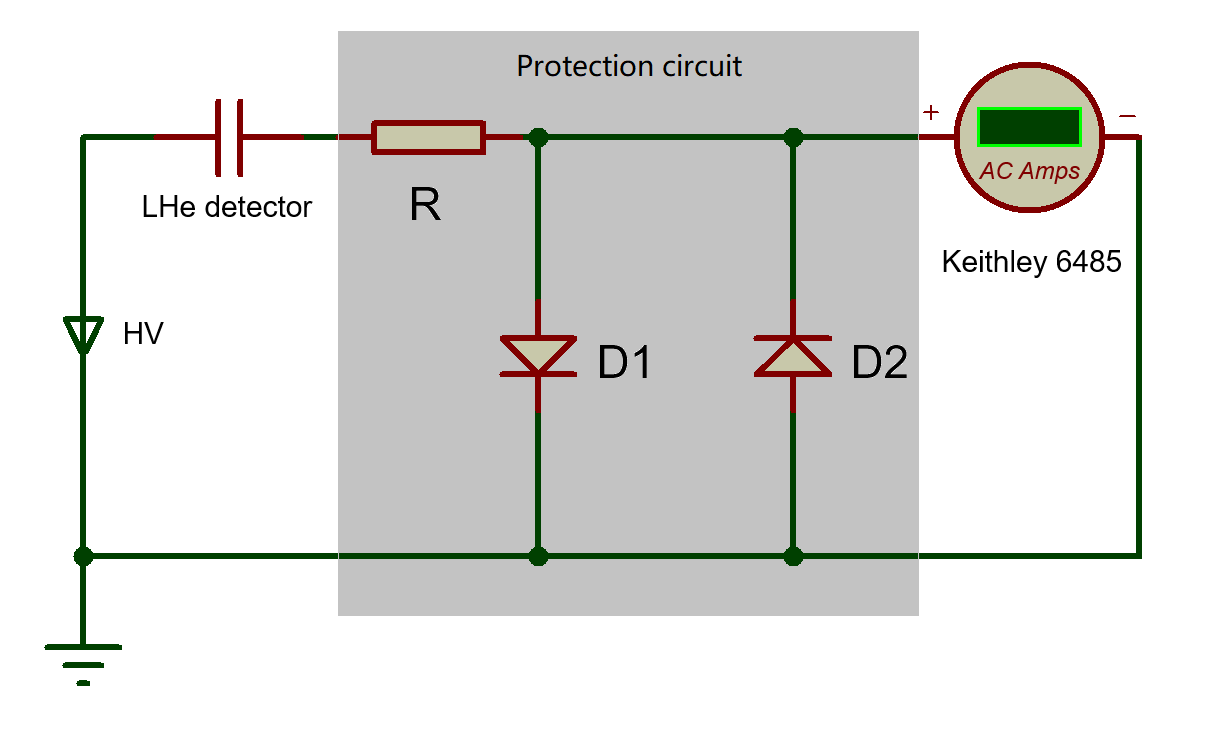}
		\caption{Fig.~\ref{fig_30gLHeDarkCurrentTest.a}. The schematic drawing of dark current tests on the 30 g LHe detector.}\label{fig_30gLHeDarkCurrentTest.a}	
		\end{subfigure}
	\quad
	\begin{subfigure}[t]{3.1in}
		\centering
		\includegraphics[scale=0.35, angle = 0]{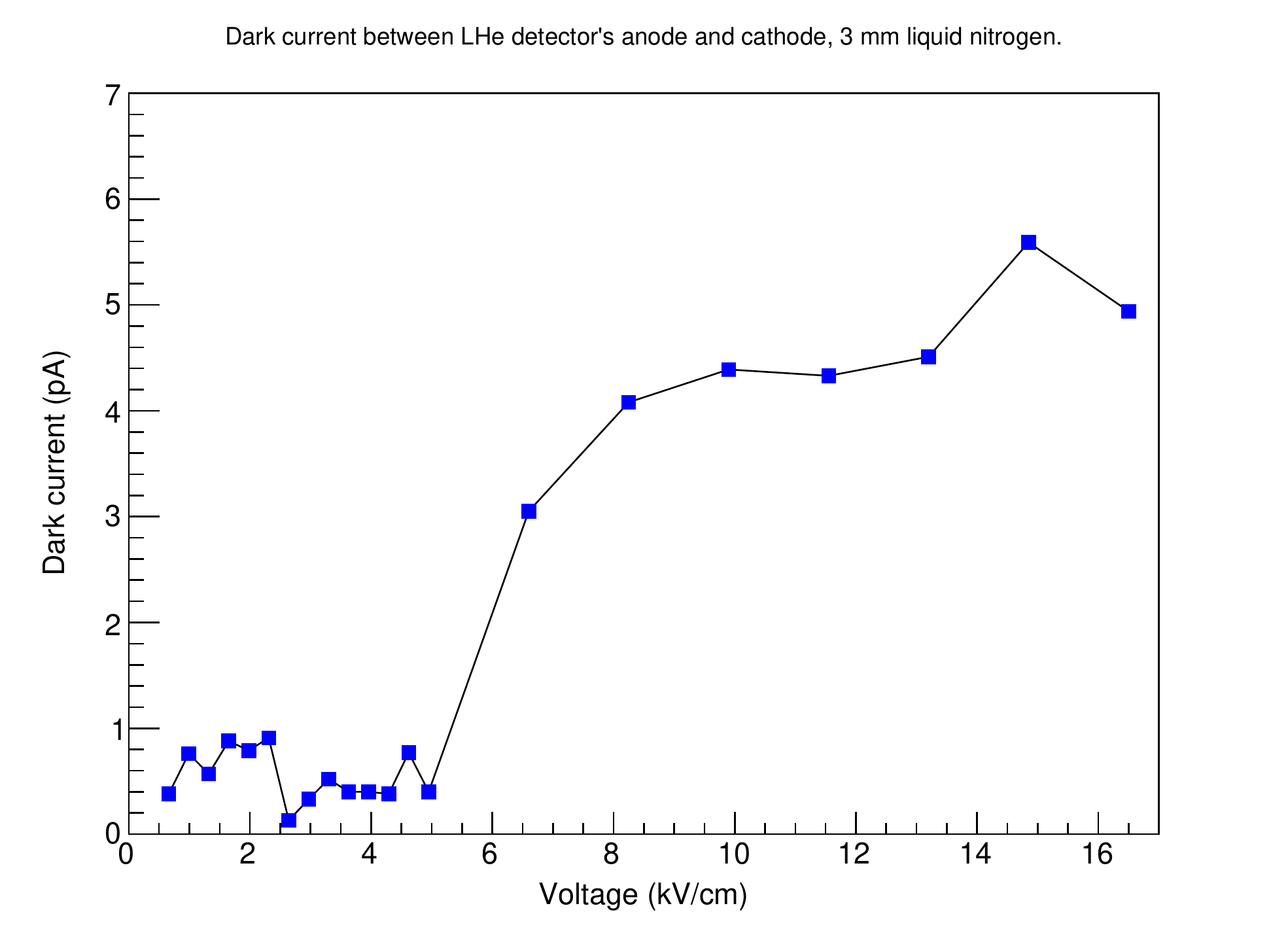}
		\caption{Fig.~\ref{fig_30gLHeDarkCurrentTest.b}. The dark current of the 30 g LHe cell, measured when filled with liquid nitrogen.}\label{fig_30gLHeDarkCurrentTest.b}
	\end{subfigure}
	\caption{Dark current test schematic drawing and measured dark current for the LHe detector filled with LN.}\label{fig_30gLHeDarkCurrentTest}
\FloatBarrier
\end{figure}

We implemented the setup as shown in Fig.~\ref{fig_30gLHeDarkCurrentTest.a} to test the dark current of the LHe system when filled with vacuum, nitrogen gas, and liquid nitrogen. The HV protection circuit in Fig.~\ref{fig_30gLHeDarkCurrentTest.a} is homemade at CIAE.  The resistor is 1 G$\Omega$, and the diodes are both 1N3595. The resistor and diodes are all low-noise elements. The picometer is Keithley 6485. For the HV power supply, we utilized ORTEC 556 and CAEN NDT1470. 

The measured dark current is $< 10 $ pA for an HV field up to 17 kV/cm when the detector was filled with three kinds of materials: vacuum, 1 atm nitrogen gas, and LN (Liquid Nitrogen)~\cite{ALETHEIA-EPJ-22}. As an example, the dark current with LN is shown in Fig~\ref{fig_30gLHeDarkCurrentTest.b}. 

\captionsetup[subfigure]{labelformat=empty}

\subsection*{The progress of TPB coating}
As demonstrated in references~\cite{McKinsey03, Ito2012, ItoSeidel13, Seidel14, Ito16, Phan20, SpiceHeRald21}, TPB (1,1,4,4-tetraphenyl-1,3-butadiene) is capable of working at LHe temperature. According to reference~\cite{Benson18},  $\sim 3~ \mu$m is the most appropriate thickness for TPB in terms of maximizing light yield. The TPB layer in DEAP-3600 detector is 3 $\mu$m~\cite{Broerman2017}, though DarkSide-50 and Darkside-20k have a much more thicker TPB layer up to $\sim 200~ \mu$m TPB~\cite{AGNES2015456, DarkSide20k17}.  \\

\begin{figure}	
	\centering
	\begin{subfigure}[t]{3.0in}
		\centering
		\includegraphics[scale=0.5]{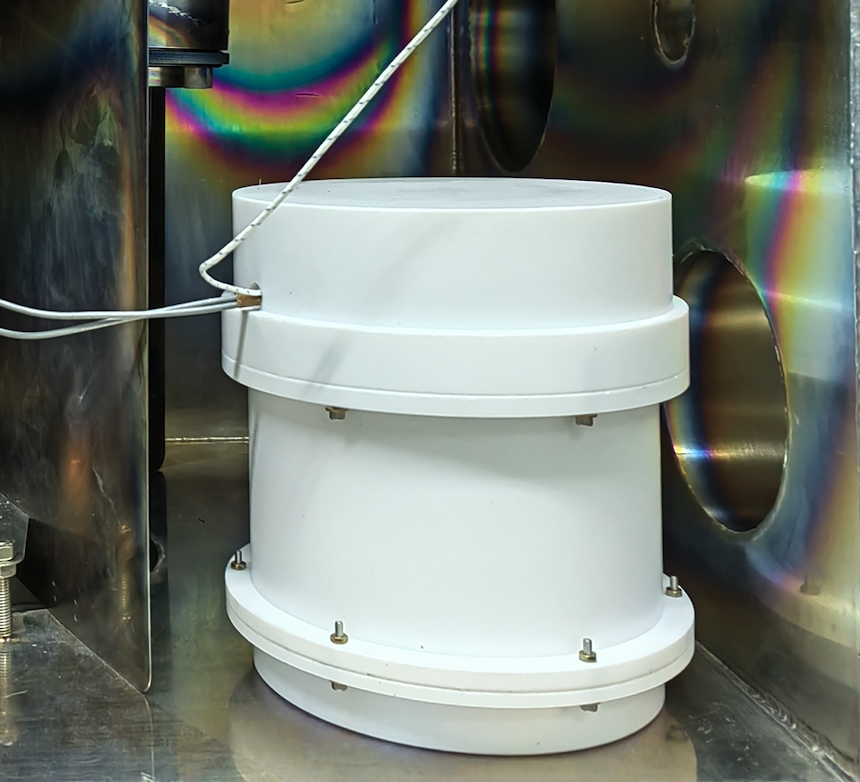}
		\caption{Fig.~\ref{fig_100gLHeCellTPBCoating.a}. We coated the white PTFE chamber shown on the plot. There are three cables on the plot, two of them are for heating, the third one is temperature sensor.} \label{fig_100gLHeCellTPBCoating.a}	
	\end{subfigure}
	\quad
	\begin{subfigure}[t]{2.8in}
		\centering
		\includegraphics[scale=0.38]{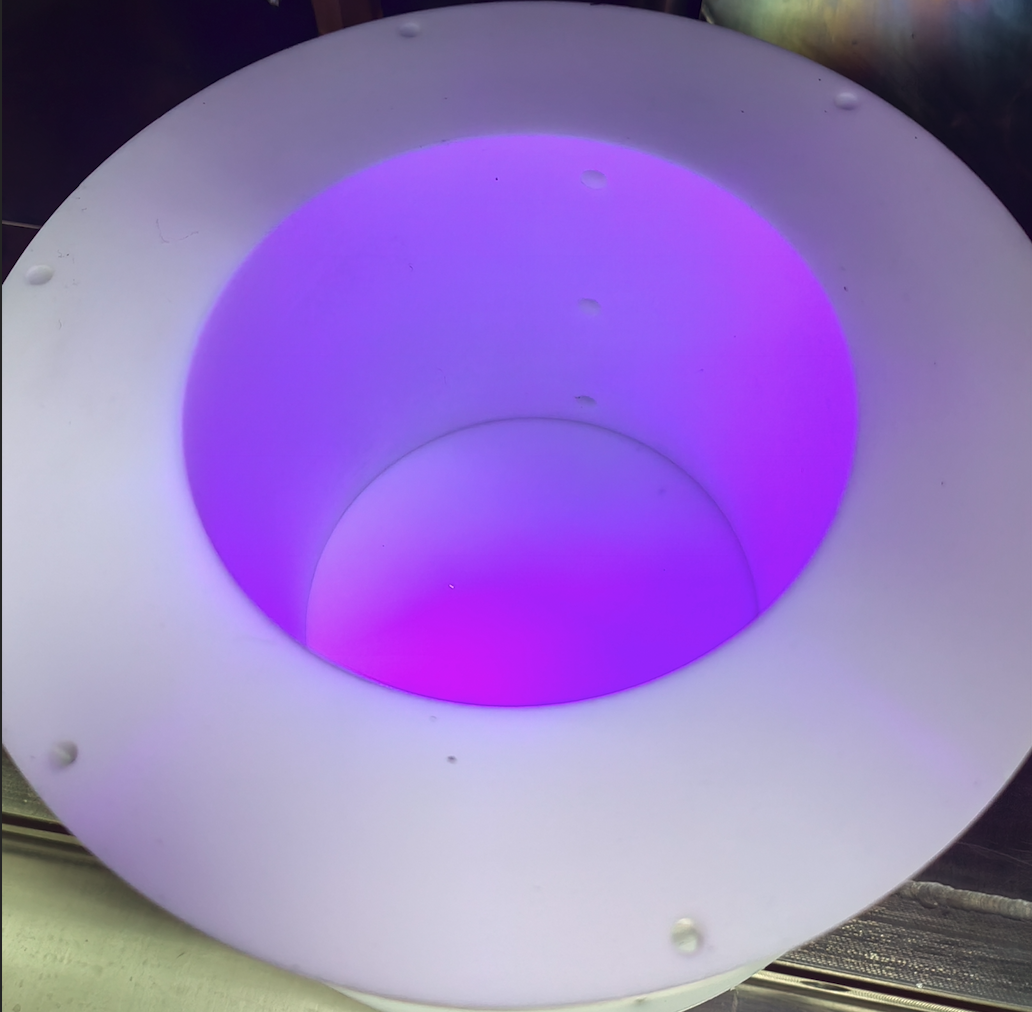}
		\caption{Fig.~\ref{fig_100gLHeCellTPBCoating.b}. $\sim 1.5~ \mu$m TPB is coated on the inner wall of the PTFE chamber. The thickness is estimated from the TPB mass evaporated.} \label{fig_100gLHeCellTPBCoating.b}
	\end{subfigure}
	\caption{$\sim 1.5~ \mu$m TPB is successfully coated on the inner wall of the PTFE chamber we designed and assembled here at CIAE.}\label{fig_100gLHeCellTPBCoating}
\FloatBarrier
\end{figure}

As a start, we coated TPB on a R = H = 10 cm cylindrical PTFE chamber as shown in Fig~\ref{fig_100gLHeCellTPBCoating.a}. We followed the coating process introduced in  references ~\cite{PollmannPhDThesis, BroermanMasterThesis, Broerman2017}. A $\sim 1.5 ~\mu$m TPB is coated on the chamber inner wall, as shown in Fig~\ref{fig_100gLHeCellTPBCoating.b}. The thickness is estimated from the TPB mass consumed during evaporation.

Later, we made two extra measures to know the thickness of the coating: (a) Adding a real-time monitor to read out the thickness of the coating online; (b) putting a few plates inside of the vessel, then comparing the mass of the plates before and after coating to figure out the TPB thickness of the plates. We coated $\sim 3.0 ~\mu$m TPB on another 10-cm size vessel. All of these three measures returned consistent thickness. We have summarized the TPB coating related work in the published paper~\cite{TPBCoutingALETHEIA22}.

\subsection*{The progress of SiPM}

Compared to high-mass WIMPs, the kinematic energy of low-mass WIMPs is smaller; accordingly, the recoil energy is minor. To detect the smaller energy, photosensors' PDE (Photon Detection Efficiency) should be as high as possible. We, therefore, choose SiPM as our photosensors thanks to its higher PDE. To implement SiPMs in ALETHEIA, the SiPM must be capable of working at $\sim$ 4 K. References~\cite{Iwai19, Cardini14} showed a specific type of Hamamatsu SiPMs can work at 6.5 K and 5 K, respectively. DarkSide-20k tested a particular type of FBK SiPMs at LAr temperature and measured the PDE greater than 40\%. However, there is no public report on FBK SiPMs can work at LHe temperature. As will introduce below, we have demonstrated that FBK SiPMs are capable of working around 4 K.

At first, we tested the IV curve at Room Temperature (RT), which is consistent with the data sheet FBK provided; for instance, the breakdown voltage is measured to be $\sim$ 30 V. For low temperatures, we mainly relied on a Gifford-McMahon (GM) cryocooler at CIAE. We successfully cooled a certain type of FBK SiPM down to 4.8 K. Since FBK company has never tested their SiPMs around 4 K, so we cannot compare with their data. However, the breakdown voltage we measured at 4.8 K and RT are 26 V and 30 V, respectively, which is consistent with what DarkSide tested at 77 K and RT: 27 V and 32 V, respectively, as shown in reference~\cite{DarkSide20k17}. 

Further, we measured the SiPM's I-V curves at 4.8 K and RT. We surprisingly found that among the total ten tested SiPMs; eight show a 10 V plateau above the breakdown voltage at a temperature between 4 K and 20 K; such an amazing plateau did not show temperatures higher than $\sim$ 20 K, even RT. We currently have no interpretation upon the feature. In addition, we also measured the SiPM with a visible light LED at $\sim 4.8$ K. For details; please refer to the paper~\cite{ALETHEIA-EPJ-22}.

\subsection*{The progress of electronics}

We have cooperated with an electronic company to design electronics suitable for our detector. Considering it might be a risk for ICs working at LHe temperature, the pre-amplifier will not work at the temperature. Instead, we developed a connection-board (no ICs) to connect SiPMs directly at 4 K, the pre-amplifier board works at RT. The two boards are connected via a 1-meter SMA cable. please refer to the paper~\cite{ALETHEIA-EPJ-22} for details. 

\section*{Summary}

DM is one of the most pressing questions to be understood and answered in fundamental physics. High-mass WIMPs detection has achieved 10$^{-48} $ cm $^2$, while still no signal has been observed yet. Low-mass WIMPs limits are only $\sim$ 10$^{-38} $ cm $^2$ , way behind high-mass WIMPs searches.

The ALETHEIA project aims to hunt for low-mass DM. Filling with the arguably cleanest material, LHe, into the arguably most competitive technology in the field, TPC, the ALETHEIA program is supposed to help answer the critical physical question today: the nature of DM. With a 1 ton*yr exposure, the detector could achieve the cross-section of DM-nucleon  to 10$^{-45} $ cm $^2$, therefore, fully touch down the $^{8}$B neutrino floor. A panel composed of world-leading physicists in the field reviewed the project with very positive comments and comprehensive suggestions. 

We have made significant progress in the past two years since the project has been officially launched in the summer of 2022.
 
(I), We built an LHe prototype detector and successfully cooled it all the way down to 4.5 K; the detector's dark current is less than 10 pA for an external HV up to 17 kV/cm. 

(II), We coated a $\sim 3.0~ \mu$m TPB layer on the inner walls of a 10-cm PTFE chamber; we developed three independent methods to figure out the coating thickness, and consistent thick results were obtained.

(III), We cooled FBK SiPMs down to 4.8 K and surprisingly found they worked pretty well; our test results are consistent with other peers'. We are the first group to observe a 10 V plateau above the breakdown voltage on FBK SiPMs at 4 - 20 K; such a plateau did not show on other temperatures.

(IV), We designed and manufactured dedicated electronics successfully to read signals out from SiPMs, both at LHe and RT temperature.

\section*{Acknowledgement}
We thank the professors who flew to Beijing in Oct 2019 to participate in the DM workshop and reviewed the project: Prof. Rick Gaitskell, Prof. Dan Hooper, Prof. Jia Liu, Prof. Dan McKinsey, Dr. Takeyaso Ito, and Prof. George Seidel. We thank Prof. Weiping Liu for helping Junhui Liao settle down at CIAE. Junhui Liao would also thank the support of the ``Yuanzhang'' funding of CIAE to launch the LHe program. We appreciate Wuxi TOFTEK Optoelectronic Technology Co., Ltd for cooperation on electronics.
 
\bibliography{ALETHEIA-snowmass-Sep2022}
\end{document}